%% file: main.tex
\documentclass[conference, 10pt]{IEEEtran}
\IEEEoverridecommandlockouts
\usepackage{cite}
\usepackage{amsmath,amssymb,amsfonts}
\usepackage{graphicx}
\usepackage{textcomp}
\usepackage{xcolor}
\usepackage{bm}
\usepackage{cases}
\usepackage[skip=1pt, belowskip=0pt]{caption}
\usepackage[ruled,vlined,linesnumbered]{algorithm2e} 
    \usepackage{setspace} 
\usepackage[makeroom]{cancel} 
\usepackage{amsthm}
\usepackage{enumitem}  

    \usepackage{physics}
    \usepackage{amsmath}
    \usepackage{tikz}
    \usepackage{mathdots}
    \usepackage{yhmath}
    \usepackage{cancel}
    \usepackage{color}
    \usepackage{siunitx}
    \usepackage{array}
    \usepackage{multirow}
    \usepackage{amssymb}
    \usepackage{gensymb}
    \usepackage{tabularx}
    \usepackage{booktabs}
    \usetikzlibrary{patterns}
    \usetikzlibrary{shapes}
    
\makeatletter
\def\BibTeX{{\rm B\kern-.05em{\sc i\kern-.025em b}\kern-.08em
    T\kern-.1667em\lower.7ex\hbox{E}\kern-.125emX}}
    
\allowdisplaybreaks
    
\include{macros_etc/macros}

\begin{document}
    
    \bstctlcite{IEEEexample:BSTcontrol}
    
    \title{Model Mismatch Trade-offs in LMMSE Estimation       
    }
    %
    %
    %
    %
    %
    %

    %
    %
    %
    %
    %
    %
    %
    %
    %
    %
    %
    %
    \author{
        \IEEEauthorblockN{
            Martin Hellkvist, 
            Ay\c ca \"Oz\c celikkale, 
            \thanks{M. Hellkvist and A.~\"Oz\c celikkale acknowledges the support from Swedish Research Council under grant 2015-04011.}
        }
        \IEEEauthorblockA{
            {Dept. of Electrical Engineering},
            {Uppsala University}, Sweden \\
            \{Martin.Hellkvist, Ayca.Ozcelikkale\}@angstrom.uu.se
        }
    }
    
    \maketitle
    
    \input{abstract}

    
    \input{macros_etc/definitions_theorems_etc}

    \input{introduction}
    \input{notation}
    \input{problem_statement}
    \input{analytical_results}
    \input{numerical_results}
    \input{conclusions}

    \input{appendix/appendix}

    \kern-0.25em
    \bibliographystyle{IEEEtran}
    \bibliography{ref}    
\end{document}

%% file: macros_etc/macros.tex
\newcommand{\Rbb}{\mathbb{R}}

\mathchardef\myhyphen="2D

\newcommand{\xhat}{\hat{\xvec}}

\newcommand{\egen}{\varepsilon}

\newcommand{\matr}[1]{\bm{#1}}
\newcommand{\Amat}{\matr{A}}

\newcommand{\Cmat}{\matr{C}}

\newcommand{\Kmat}{\matr{K}}

\newcommand{\Wmat}{\matr{W}}

\newcommand{\avec}{\matr{a}}

\newcommand{\vvec}{\matr{v}}

\newcommand{\wvec}{\matr{w}}
\newcommand{\xvec}{\matr{x}}
\newcommand{\yvec}{\matr{y}}
\newcommand{\zvec}{\matr{z}}

\newcommand{\Imat}{\matr{I}}

\newcommand{\eye}[1]{\Imat_{#1}}

\DeclareMathOperator{\Ebb}{\mathbb{E}}
\newcommand{\eunder}[1]{\underset{#1}{\Ebb}}

\DeclareMathAlphabet{\mymathbb}{U}{BOONDOX-ds}{m}{n}

\newcommand{\T}{^\mathrm{T}}

\newcommand{\p}{^+}

\let\oldin\in
\renewcommand{\in}{{\,\oldin\,}}

\let\oldnotin\notin
\renewcommand{\notin}{{\,\oldnotin\,}}

\renewcommand{\th}{\textsuperscript{th}}

\renewcommand{\tr}{\operatorname{tr}}

\long\def\/*#1*/{}

\newcommand{\Nc }{\mathcal{N}}

\definecolor{myblue}{rgb}{0.3328, 0.3539, 0.7758}
\definecolor{myblue2}{rgb}{0.0328, 0.0539, 0.4758}
\definecolor{mygreen2}{rgb}{ 0.0328 0.4758 0.0539} 
\definecolor{mygreen3}{rgb}{ 0.0328 0.1758 0.0539} 
\definecolor{myred}{rgb}{0.4758, 0.0328, 0.0539}
\definecolor{myred2}{rgb}{0.75, 0.0328, 0.0539}

\newcommand{\Khat}{\hat{\Kmat}}
\newcommand{\sigmahat}{\hat{\sigma}}

%% file: abstract.tex
\begin{abstract}
    We consider a linear minimum mean squared error (LMMSE) estimation  framework with model mismatch 
    where the assumed model order is smaller than that of the underlying linear system which generates the data used in the estimation process.
    By modelling the regressors of the underlying system as random variables,
    we analyze the average behaviour of the mean squared error (MSE).
    Our results quantify how the MSE depends on the interplay between the number of samples and the number of parameters in the underlying system and in the assumed model. 
    In particular, 
    if the number of samples is not sufficiently large,
    neither increasing the number of samples nor the assumed model complexity is sufficient to guarantee a performance improvement.
\end{abstract}

%% file: macros_etc/definitions_theorems_etc.tex
\newtheorem{thm}{\bf{Theorem}}
\newtheorem{cor}{\bf{Corollary}}
\newtheorem{lem}{\bf{Lemma}}
\newtheorem{prop}{\bf{Proposition}}
\newtheorem{rem}{\bf{Remark}}

\theoremstyle{remark} 
\newtheorem{defn}{\bf{Definition}}[section]
\newtheorem{ex}{\bf{Example}}[section]
\newtheorem{myexp}{\bf{Experiment}}

\newenvironment{theorem}
{\par\noindent \thm \begin{itshape}\noindent}
{\end{itshape}}

\newenvironment{lemma}
{ \par\noindent  \lem \begin{itshape}\noindent}
{\end{itshape}
}

\newenvironment{corollary}
{\par\noindent  \cor \begin{itshape}\noindent}
{\end{itshape}
}

\newenvironment{remark}
{\par\noindent \rem \begin{itshape}\noindent}
{\end{itshape}
}

\newenvironment{definition}
{\par\noindent \defn \begin{itshape}\noindent}
{\end{itshape}}

\newenvironment{experiment}
{\vspace{3pt} \par\noindent \myexp \begin{itshape} \noindent}
{\end{itshape}\vspace{6pt}}

\newenvironment{example}
{\vspace{2pt} \par\noindent \ex} 
{\vspace{2pt}}

%% file: introduction.tex
\section{Introduction}
Described in several text books and used in a wide range of applications,
the linear minimum mean square error (LMMSE) estimator 
\cite{kailath_sayed_hassibi_linear_estimation_2000, kay1993fundamentals} 
is one of the fundamental estimation methods of signal processing.
Being a Bayesian estimation approach,
the parameters of interest are modeled as random variables with some joint probability density function (pdf),
based on some background knowledge.
%
The LMMSE estimator is the optimal estimator out of all the possible linear (more precisely affine) estimators in terms of minimizing the mean squared error (MSE),
and it only depends on the mean and covariances.  
If the assumed covariance matrices are  inaccurate,
which is generally the case for real-world problems,
then the performance of the computed LMMSE estimator can be 
suboptimal.
%
In this article, we focus on characterizing such performance degradation. 


We consider the underlying system $\yvec = \Amat \xvec  + \vvec$,
where $\yvec$ is the observed output vector, 
$\Amat$ is the matrix of regressors, 
$\vvec$ is some unknown noise vector and the vector
$\xvec$ denotes the unknown model parameters which we want to estimate.
We model $\xvec$ and $\vvec$ as random vectors,
and propose an LMMSE estimation framework which allows us to systematically study the MSE when only a subset $\Amat_S$ of the columns in $\Amat$ are available for estimation.
%
In particular, the mismatched estimator is based on the assumed system $\yvec = \Amat_S \xvec_S  +  \zvec$,
where the assumed number of unknowns (length of $\xvec_S$) is smaller than the number of unknowns in the underlying system (length of $\xvec$).
%
We model the regressors in $\Amat$ as random variables
and derive an analytical expression for the expected MSE of the low order LMMSE estimator, over the distribution of $\Amat$.

A range of methods have been proposed for robustness against uncertainties or model flaws in the LMMSE estimation.
Methods to deal with covariance matrix uncertainties have been presented in \cite{
    lederman_tabrikian_constrained_2006, 
    mittelman_miller_robust_2010, Zachariah_Shariati_Bengtsson_Estimation_2014},
and in \cite{Liu_Zachariah_Stoica_Robust_2020} the effect of having missing features, i.e., unknowns, in the underlying model was investigated.
Robustness have been also investigated under a classical estimation framework where the unknown is modelled as deterministic, such as for uncertainties in the regressors \cite{eldar_bental_nemirovski_robust_2005}.
Further model mismatch trade-offs in classical estimation settings have  been studied, focusing on the relationship between model size and number of observations  \cite{breiman_how_nodate, belkin_two_2019}.
In our setup,
only parts of the regressors are available for estimation,
and the respective models on $\xvec_S$ and $\wvec$ do not match the underlying models $\xvec$ and $\vvec$, constituting a hybrid setting with uncertain regressors 
and a model mismatch in the unknowns and the noise.
We study the average MSE performance under a model mismatch and with an isotropic Gaussian model on the regressors. 
Our contributions can be summarized as follows: 
i) Our analytical results show that the MSE depends on
the respective signal powers of $\xvec$, $\xvec_S$ and $\vvec$, but not 
 on the general covariance structure of the unknowns $\xvec$. ii) 
These results quantify how the MSE heavily depends on the relation between the number of samples,
the underlying and the assumed model orders: 
If the number of samples is not sufficiently large,
then the performance is not guaranteed to improve by increasing the number of samples or the assumed model complexity.
In particular, 
lowering the assumed model order can improve the performance even when the number of samples is larger than the number of unknowns in the underlying system.

The rest of the paper is organized as follows:
In Section~\ref{sec:problem},
we provide the problem formulation.
In Section~\ref{sec:analytical}, we present and discuss our main analytical results,
which are numerically verified in Section~\ref{sec:numerical}.
Conclusions are summarized in Section~\ref{sec:conclusions}.


%% file: notation.tex
\textbf{Notation:}
        We denote the Moore-Penrose pseudoinverse and the transpose of a matrix $\Amat$ as $\Amat\p$ and $\Amat\T$, respectively.
        The $p\times p$ identity matrix is denoted as $\eye{p}$.
        The Euclidean norm and trace operator are denoted by $\|\cdot\|$ and  $\tr(\cdot)$, respectively.
        We use the notation $\eunder{\xvec}$ or $\Ebb_{\xvec}$ to emphasize that the expectation is taken with respect to the random variable $\xvec$.
        For two column vectors
        $\zvec$, $\wvec$, we denote their covariance matrix by $\Kmat_{\zvec\wvec} = \Ebb_{\zvec,\wvec}[(\zvec - \Ebb_{\zvec}[\zvec]) (\wvec-\Ebb_{\wvec}[\wvec])\T]$.
        For auto-covariance matrices, 
        we write the subscript only once: $\Kmat_{\zvec} = \Kmat_{\zvec\zvec}$.
      

%% file: problem_statement.tex
\kern2em
\section{Problem Statement}\label{sec:problem}
\kern-0.5em
\subsection{The Underlying System} 
\kern-0.5em
The observations 
$\yvec$ come from the following linear system
\begin{equation}\label{eqn:model}
    \yvec = \Amat\xvec + \vvec,
\end{equation}
where  $\xvec = [x_1,\, \dots,\, x_n]\T \in\Rbb^{p\times 1}$ denotes the unknowns,
$\yvec = [y_1,\, \dots,\, y_n]\T \in \Rbb^{n\times 1}$ denotes the vector of observations,
$\Amat = [\avec_1\, | \, \cdots \, | \, \avec_n]\T\in \Rbb^{n\times p}$ denotes the known matrix of regressors, 
and 
$\vvec = [v_1,\,\dots,\, v_n]\in\Rbb^{n\times 1}$ denotes the unknown noise.
Here, $\xvec$ and $\vvec$ are modeled as zero-mean uncorrelated random vectors.
Note that $\avec_i\T$ denotes the $i$\th{} row of $\Amat$, 
i.e., the  regressors corresponding to the observation $y_i = \avec_i\T \xvec + v_i$.

Consider the class of linear estimators,
i.e.,
the estimators such that the estimate $\xhat$ is a linear function of the vector of observations $\yvec$,
with $\xhat = \Wmat \yvec$ where $\Wmat\in\Rbb^{p\times n}$.   
The mean squared error (MSE) associated with $\Wmat$ is given by
\begin{equation}\label{eqn:mse}
    J(\Wmat) 
    = \eunder{\xvec, \yvec}\left[\|\xvec - \xhat\|^2\right]
    = \eunder{\xvec, \yvec}\left[\|\xvec - \Wmat \yvec\|^2\right].
\end{equation}
Under the linear model in \eqref{eqn:model},  $J(\Wmat)$ is found as
\begin{align}
    J(\Wmat) 
    &= \eunder{\xvec, \yvec}\left[\|\xvec - \Wmat(\Amat\xvec +  \vvec)\|^2\right] \\
    &= \eunder{\xvec, \vvec}\left[\|(\eye{p} - \Wmat\Amat) \xvec - \Wmat \vvec\|^2\right] \\
    &= \tr( (\eye{p} \! - \! \Wmat\!\Amat) \Kmat_{\xvec} (\eye{p} \! - \! \Wmat \! \Amat)\T \! + \! \Wmat \Kmat_{\vvec} \Wmat\T ).
\end{align}

The linear minimum MSE (LMMSE) estimator, 
i.e., the matrix $\Wmat$ that minimizes the MSE $J(\Wmat)$ over all $\Wmat\in\Rbb^{p\times n}$, is  
given by \cite{kailath_sayed_hassibi_linear_estimation_2000} 
\begin{align}
    \Wmat_O &=\Kmat_{\xvec \yvec} \Kmat_{\yvec}\p 
    = \Kmat_{\xvec} \Amat\T ( \Amat \Kmat_{\xvec} \Amat\T + \Kmat_{\vvec} )\p, \label{eqn:W:lmmse_true_A}
\end{align}
where we have used the fact that under \eqref{eqn:model}, we have
$\Kmat_{\xvec \yvec} =\Kmat_{\xvec} \Amat\T$ and $\Kmat_{\yvec} =\Amat \Kmat_{\xvec} \Amat\T + \Kmat_{\vvec}$.
In \eqref{eqn:W:lmmse_true_A} we have used the Moore-Penrose pseudoinverse, 
rather than the ordinary inverse,
which, as discussed in \cite[Theorem 3.2.3]{kailath_sayed_hassibi_linear_estimation_2000} will minimize the MSE regardless of whether $\Kmat_{\yvec}$ is  singular or not.

\kern-0.75em
\subsection{Model Mismatch and Assumed Model}\label{sec:lmmse_partial}
\kern-0.5em
In this paper, our focus is on estimation under a model mismatch.
In particular, 
we consider the case that the LMMSE estimator relies on an incorrect signal model such that i) only a subset of the unknowns $\xvec_i$ are assumed to be present in the system equation; ii) the assumed covariances are possibly inconsistent with the underlying system in \eqref{eqn:model}.   
Let this subset of $\xvec$ be denoted by 
$\xvec_S\in\Rbb^{p_S \times 1}$ and its complement 
(i.e., the elements of $\xvec$ that are not in $\xvec_S$) by $\xvec_C\in\Rbb^{p_C \times 1}$,
where  $p = p_S + p_C$.
Let $\Amat_S \in \Rbb^{n\times p_S}$ and $\Amat_C \in \Rbb^{n \times p_C}$ denote the submatrices of $\Amat$ consisting of the columns corresponding to the indices that are in $\xvec_S$ and in $\xvec_C$, respectively. 

The estimator uses the following partial model
\setlength{\abovedisplayskip}{2pt}
\setlength{\belowdisplayskip}{2pt}
\begin{equation}\label{eqn:model_assumed}
    \yvec = \Amat_S \xvec_S + \zvec,
\end{equation}
where $\xvec_S$ and the noise $\zvec$ are assumed to be uncorrelated
and zero-mean,
and $\Amat_S$ is known. 
Here, the respective \textit{assumed} covariance matrices for $\xvec_S$ and the noise $\zvec$ are given by $\Khat_{\xvec_S}$ and $\Khat_z$.  
We have used the notation $\Khat$ to emphasize that these covariance matrices are not necessarily the same as the ones that can be derived from \eqref{eqn:model}.
Hence, there is a model mismatch between \eqref{eqn:model_assumed} and \eqref{eqn:model}.
According to \eqref{eqn:model_assumed}, other covariance matrices of interest are given by
\begin{equation}
    \Khat_{\xvec_S \yvec} \!
    = \!\!\!\! \eunder{\xvec_S, \yvec}\!\left[\xvec_S \yvec\T\right]  \!
    = \!\!\!\! \eunder{\xvec_S, \zvec}\!\left[\xvec_S (\Amat_S \xvec_S \! + \! \zvec)\T\right]  \!
    = \! \Khat_{\xvec_S} \Amat_S\T,
\end{equation}
\begin{equation}
    \Khat_{\yvec} \!
    = \!\!\!\! \eunder{\xvec_S, \zvec}\!\!\left[(\Amat_S \xvec_S \! + \! \zvec) (\Amat_S \xvec_S \! + \! \zvec)\T\right] 
    \!\! = \! \Amat_S \Khat_{\xvec_S} \Amat_S\T  \! + \! \Khat_{\zvec}. 
\end{equation}

Let $\xhat_S =\Wmat_S \yvec$ be an estimate of $\xvec_S$,
where $\Wmat_S\in\Rbb^{p_S\times n}$.
Then the corresponding MSE for $\xvec_S$ is given by
\begin{equation}\label{eqn:mse_S_xhat_S}
    J_S(\Wmat_S) 
    = \!\! \eunder{\xvec_S, \yvec} \left[\|\xvec_S \! - \! \xhat_S\|^2\right]
    = \!\! \eunder{\xvec_S, \yvec} \left[\|\xvec_S \! - \! \Wmat_S \yvec\|^2\right].
\end{equation}
\setlength{\abovedisplayskip}{4pt}
\setlength{\belowdisplayskip}{4pt}
An explicit expression for $J_S(\Wmat_S)$ is provided in \eqref{eqn:MSE_trace_xhat_S}.
The corresponding LMMSE estimator, 
assuming \eqref{eqn:model_assumed}, 
is given by 
\begin{equation}\label{eqn:lmmse_partial}
   \!\xhat_{\!S} \!\!
    = \!\! \Wmat_{\!\!S} \yvec \! 
    = \! \Khat_{\xvec_S \! \yvec} \Khat_{\!\yvec}\p \! \yvec \!
    = \!\! \Khat_{\xvec_S}\! \Amat_{\!S}\T(\!\Amat_{\!S} \Khat_{\xvec_S} \Amat_{\!S}\T \!+ \! \Khat_{\zvec})\p \!\yvec  ,
\end{equation}

We note that the estimator in \eqref{eqn:lmmse_partial} would be the true LMMSE estimator if the observations $\yvec$ were in fact generated by the model in \eqref{eqn:model_assumed}.
However, this is not the case. 
Here, 
$\yvec$ actually comes from the underlying system in \eqref{eqn:model},
hence the true LMMSE estimate of $\xvec_S$,
minimizing the MSE in \eqref{eqn:mse_S_xhat_S}, is
\begin{equation}\label{eqn:lmmse_partial:correct}
    \xhat_S = \Kmat_{\xvec_S} \Amat_S\T (\Amat \Kmat_{\xvec} \Amat\T + \Kmat_{\vvec})\p \yvec.
\end{equation}
To summarize our setting,
$\yvec$ is generated by the system in \eqref{eqn:model},
while the estimation is performed under the assumption that $\yvec$ is generated by \eqref{eqn:model_assumed}. 
Hence, 
the LMMSE estimator in \eqref{eqn:lmmse_partial} is used instead of the correct estimator in \eqref{eqn:lmmse_partial:correct}.
In other words, we consider LMMSE estimation under a model mismatch.

In order to take into account the part of $\xvec$ that is not estimated in this partial setting, 
i.e., $\xvec_C$, 
we also define the MSE associated with the whole vector $\xvec$ under $\Wmat_S$ as
\begin{equation}
    J(\Wmat_S) = J_S(\Wmat_S) + \tr(\Kmat_{\xvec_C}).
\end{equation}
Note that the subscript $S$ in $J_S(\cdot)$ emphasizes that the error is over $\xvec_S$ whereas $J(\cdot)$ refers to the error in the whole vector $\xvec$.
Here, $J(\Wmat_S)$ corresponds to the error associated with estimating $\xvec_S$ with $\Wmat_S$ while setting the estimate of $\xvec_C$ to $E[\xvec_C]=0$. 

\subsection{Expected MSE over Regressors}\label{sec:mse_random_A}
\label{sec:problem_expected_MSE}
\kern-0.25em
We are interested in the average behaviour of the MSE of the partial LMMSE estimator in \eqref{eqn:lmmse_partial}
over regressor matrices  $\Amat$.
We model $\avec_i$'s as independent and identically distributed (i.i.d.) Gaussian random vectors,
i.e.,
$\avec_i \! \sim \! \Nc(0, \Kmat_{\avec})$, $\forall\,i$ with $\Kmat_{\avec} =\eye{p}$. 
The \textit{expected MSE}  over the distribution of $\Amat$'s is given by 
\begin{equation}\label{eqn:full_mse_partial_lmmse_S}
    \egen_S(p_S, n) 
    = \eunder{\Amat} \left[ J_S(\Wmat_S) \right] 
    = \eunder{\Amat} \left[ J_S(\Khat_{\xvec_S \yvec} \Khat_y\p) \right].
\end{equation}
Note that this is the expected MSE associated with $\xvec_S$. 
Here $\Wmat_S$ is a function of $\Amat$
(more precisely a function of $\Amat_S$, a submatrix of $\Amat$),
and $\yvec$ varies with $\Amat$. 
We are interested in how the MSE varies for different choices of $p_S$, i.e., the number of estimated parameters,
and $n$, i.e., the number of samples in $\yvec$. Hence, $ \egen_S(p_S, n)$ is defined as a function of these variables. 

Here, we analyze the MSE from the perspective of repeated experiments using different matrices $\Amat$,  
hence we here model $\Amat$ as a random matrix. 
Nevertheless, 
note that while doing the LMMSE estimation, 
$\Amat$ and $\Amat_S$ are known in \eqref{eqn:W:lmmse_true_A} and \eqref{eqn:lmmse_partial}, 
respectively. 

We similarly define the expected MSE associated with the whole vector $\xvec$ as
\begin{equation}\label{eqn:full_mse_partial_lmmse}
    \egen(p_S, n) = \eunder{\Amat}[J(\Wmat_S)] = \egen_S(p_S, n) + \tr(\Kmat_{\xvec_C}),
\end{equation}
which in addition to \eqref{eqn:full_mse_partial_lmmse_S} takes into account the power of the signal $\xvec_C$ that is disregarded by the assumed model \eqref{eqn:model_assumed}.

As a part of our analysis of  $\varepsilon(p_S,n)$,
we also compare it to the expected MSE associated with the full LMMSE estimator
\begin{equation}
    \Bar{\varepsilon} = \eunder{\Amat}[J(\Wmat_O)] = \eunder{\Amat}[J(\Kmat_{\xvec\yvec}\Kmat_{\yvec}\p)].
\end{equation}
where $\Wmat_O$ is the estimator in \eqref{eqn:W:lmmse_true_A}.

%% file: analytical_results.tex
\section{Expected MSE under a Model Mismatch}
\label{sec:analytical}
\vspace*{-2pt}

We now provide an explicit expression $J_S( W_S)$ of \eqref{eqn:mse_S_xhat_S}. 
Plugging in $y$ from the underlying model in \eqref{eqn:model}, 
\begin{align}\label{eqn:MSE_trace_xhat_S}
    \begin{split}
    J_S&( \Wmat_S) 
    =  \eunder{\xvec, \vvec}\big[\|\xvec_S - \Wmat_S(\Amat_S\xvec_S + \Amat_C \xvec_C + \vvec)\|^2\big]
    \end{split}\\
    \begin{split}
        &=  \eunder{\xvec, \vvec}\big[\|(\eye{p_S} \! - \! \Wmat_S \Amat_S) \xvec_S
        \! - \! \Wmat_S \Amat_C \xvec_C - \Wmat_S \vvec\|^2 \big]
    \end{split}\\
    \begin{split}
        & = \tr\big((\eye{p_S} \! - \! \Wmat_S \Amat_S) \Kmat_{\xvec_S} (\eye{p_S} - \Wmat_S \Amat_S)\T \\
        & \quad\quad + \Wmat_S \Amat_C \Kmat_{\xvec_C} \Amat_C\T \Wmat_S\T + \Wmat_S \Kmat_{\vvec} \Wmat_S\T  \\
        & \quad\quad - 2 \Wmat_S \Amat_C \Kmat_{\xvec_C \xvec_S} (\eye{p_S} - \Wmat_S \Amat_S)\T \big).
    \end{split}
\end{align}

The following result describes the generalization error associated with the partial LMMSE estimator in \eqref{eqn:lmmse_partial}:
\begin{theorem}\label{thm:mse}
    With $\Kmat_{\avec} = \eye{p}$, $\Khat_{\xvec_S} = \eye{p_S}$ and $\Khat_{\zvec} = 0$,
    i.e., the noise $\zvec$ is assumed to be zero,
    the partial LMMSE estimator in \eqref{eqn:lmmse_partial} has the expected MSE
    \begin{align} \label{eqn:thm:digaonal:mse}
    \begin{split}
        \egen_S(p_S, n) 
        =  
        \tr(\Kmat_{\xvec_S})
        \bigg( 
        1 & - \frac{\min\{p_S, n\}}{p_S} 
        \bigg) \\ 
         + \gamma & \left(\tr(\Kmat_{\xvec_C}) + \frac{1}{n} \tr(\Kmat_v)\right),
    \end{split}
    \end{align}
    with $\gamma$ defined as
    \begin{subnumcases}{\label{eqn:gamma}\hspace{-10pt} \gamma=}
        \tfrac{p_S}{n - p_S - 1 } & for $p_S < n - 1,$ \label{eqn:gamma_a}\\
        \tfrac{n}{p_S - n - 1 } & for $p_S > n + 1,$ \label{eqn:gamma_b}\\
        +\infty & otherwise, \label{eqn:gamma_c}
    \end{subnumcases}
    where $p = p_S + p_C$.
\end{theorem}

\noindent Proof: See Section~\ref{proof:thm:mse}.

Theorem~\ref{thm:mse} quantifies the dependence of the  expected MSE $\egen_S$  on the individual powers of  $\xvec_S$, $\xvec_C$ and the noise $\vvec$,
i.e., $\tr(\Kmat_{\xvec_S})$, $\tr(\Kmat_{\xvec_C})$ and $\tr(\Kmat_{\vvec})$.
It also reveals that the error does not depend on the covariance between $\xvec_S$ and $\xvec_C$,
or on the general structure of $\Kmat_{\vvec}$.

{\bf Effect of $\gamma$:}
The factor $\gamma$, hence  $\egen_S$,
can take extremely large values if the number of samples $n$ is too close to the number of estimated parameters $p_S$.
We observe that if both $\xvec_C$ and $\vvec$ are identically zero,
then $\gamma$ does not affect the MSE,
however this is generally not the case.

We continue the discussion of the behaviour of $\egen_S$ by considering the following scenarios of $n$ versus $p_S$:
\textbf{i)} $n > p_S$, and \textbf{ii)} $n < p_S$.

{\textbf{i) $n>p_S$}}:
Here, the MSE component from $\tr(\Kmat_{\xvec_S})$ is constantly zero,
and $\gamma$ in \eqref{eqn:gamma_a} decreases monotonically with an increasing $n$.
Hence, if the noise level per sample does not increase with the number of samples, 
i.e., if $\tr(\Kmat_{\vvec})/n$ doesn't increase with $n$,
then the MSE monotonically decreases with increasing $n$. 
Regarding the MSE's dependency on $p_S$,
we will show in Corollary~\ref{col:n_SNR_dependency}
that if $n$ is not large enough,
then the expected MSE is not guaranteed to improve with $p_S$, under some additional constraints.

{\textbf{ii) $n < p_S$}}:
The result in Theorem~\ref{thm:mse} shows that 
the performance is not guaranteed to improve by having more samples.
In \eqref{eqn:gamma_b},
we see that for $n < p_S$, 
$\gamma$ increases with $n$,
hence the expected MSE can also increase with $n$.
In particular, 
as we will illustrate with numerical examples in Section~\ref{sec:numerical},
the power in $\xvec_S$ must be significantly larger than the combined powers of $\xvec_C$ and $\vvec$,
in order for the MSE to decrease as $n$ increases.
For such small $n$, it is also not immediately apparent which choice of $p_S$ gives the lowest MSE.
This insight is illustrated in the numerical examples in  Section~\ref{sec:numerical}.

The following corollary is a special case of Theorem~\ref{thm:mse} where the powers in $\xvec_S$ and $\xvec_C$ are directly proportional to $p_S$ and $p_C$ and the noise level per sample  is constant:
\begin{corollary}\label{col:mse}
    Consider the setting of Theorem~\ref{thm:mse}, 
    with $\tr(\Kmat_{\xvec_S}) = \sigma_x^2 \, p_S > 0$, $\tr(\Kmat_{\xvec_C})= \sigma_x^2 \, p_C \geq 0$
    and $\tr(\Kmat_{\vvec}) = n \sigma_v^2 > 0$,
    then the partial LMMSE estimator in \eqref{eqn:lmmse_partial} has the following expected MSE:
    \begin{align}
        \egen_S(p_S, n) 
        =  \sigma_x^2 (p_S - \min\{p_S, n\})  + \gamma( \sigma_x^2\, p_C + \sigma_v^2).
    \end{align}
    
\end{corollary}

\noindent Proof: This result is readily obtained by plugging in the respective assumptions on $\Kmat_{\xvec}, \Kmat_{\xvec_S}, \Kmat_{\xvec_C}$ and $\Kmat_{\vvec}$ into \eqref{eqn:thm:digaonal:mse}.

Under the given additional assumptions, Corollary~\ref{col:mse} gives a clear characterization of the dependence of the MSE on the respective dimensions of $\xvec_S$ and $\xvec_C$, 
the number of samples $n$,
and the power levels $\sigma_x^2$ and $\sigma_v^2$.

While our Bayesian problem formulation is different than the classical estimation  setting of \cite{belkin_two_2019},
the result in this corollary describes the same phenomenon as studied in \cite[Theorem 2.1]{belkin_two_2019}.
In \cite{belkin_two_2019},  according to least-squares estimation setting \cite[Ch.8]{kay1993fundamentals},
the performance is measured by the residuals $y_i - \avec_i\T\xhat$,
i.e., the error made when predicting $\yvec$ with $\xhat$. 
On the other hand, in this paper we focus on the error associated with the estimate $\xhat$. Nevertheless, 
the expected error for the estimate of a new $y_j$ satisfies
$\Ebb_{y_j,\xvec,\xhat,\avec_j,\vvec}[(y_j - \avec_j\T\xhat)^2] =\Ebb_{\xvec,\xhat}[\|\xvec- \xhat\|^2]+\sigma_v^2$,
with $\avec_j\sim\Nc(0,I_p)$.   
Moreover, the least-squares (LS) estimator in \cite{belkin_two_2019} matches our estimator under the assumptions of Theorem~\ref{thm:mse},
i.e., $\xhat=\Amat_S\p \yvec$. 

\begin{corollary}\label{col:n_SNR_dependency}
    Consider the setting of Corollary~\ref{col:mse}.
    Let $n > p + 1$.
    Then the expected MSE $\egen(p_S,n)$ decreases monotonically with $p_S$ if 
    \begin{equation}
        n > p + \sigma_v^2/\sigma_x^2 + 1.
    \end{equation}
    Furthermore, $\egen(p_S,n)$ increases monotonically with $p_S$ if 
    \begin{equation}
        n < p + \sigma_v^2/\sigma_x^2 + 1.
    \end{equation}
    
\end{corollary}

Proof: This result can be found by treating $p_S$ as a continuous variable and taking the derivative of $\egen(p_S, n) = \egen_S(p_S,n) + \tr(K_{x_C})$ w.r.t. $p_S$, and solving the inequalities $\partial \egen/\partial p_S < 0$ and 
$\partial \egen/\partial p_S > 0$, for n.

Corollary~\ref{col:n_SNR_dependency} shows that $n$ needs to be sufficiently large to guarantee a performance gain with an increase in the assumed model's complexity $p_S$.
It also shows that for $p+1 < n < p + \sigma_v^2/\sigma_x^2 + 1$,
the MSE increases with $p_S$.
We note that the bound is larger for worse signal-to-noise (SNR) ratios $\sigma_x^2/\sigma_v^2$,
increasing as the SNR decreases.

%% file: numerical_results.tex
\kern-0.25em
\section{Numerical Results}\label{sec:numerical}
\kern-0.125em
\begin{figure}[t]
    \centering
    \includegraphics[scale=0.9]{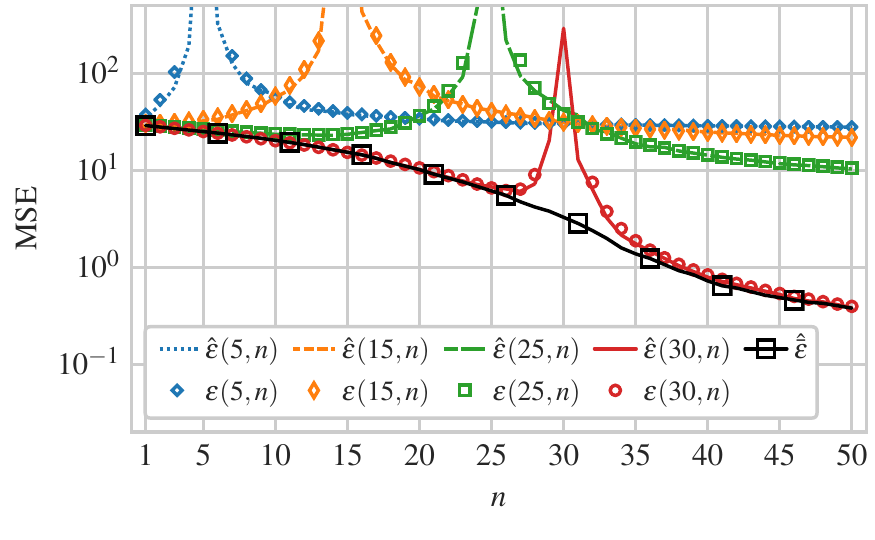}
    \caption{
        Empirical and analytical MSE versus the number of samples,
        for the partial LMMSE estimator in the high SNR case \textit{(S1)}.
    }
    \label{fig:exp1b}
\end{figure}

\begin{figure}[t]
    \centering
    \includegraphics[scale=0.9]{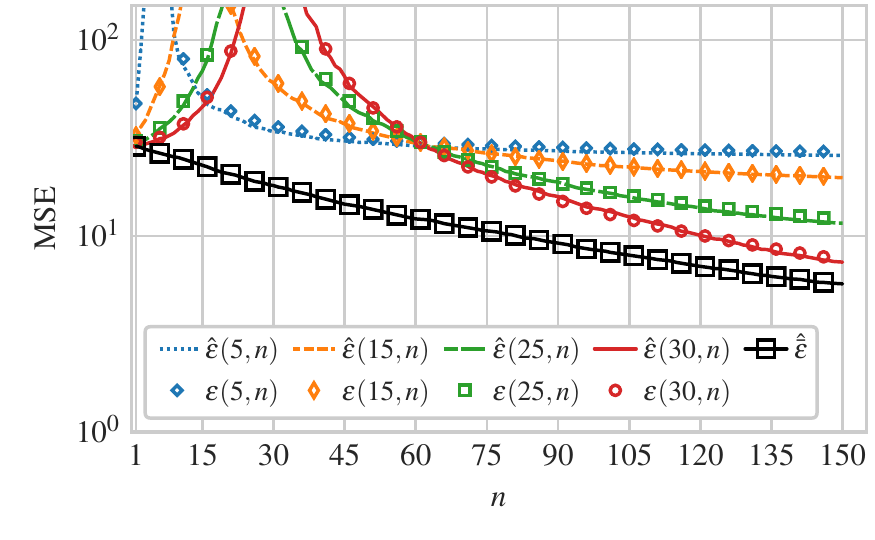}
    \caption{
       The MSE in the low SNR case \textit{(S2)}.
    }
    \vspace{-1.15em} 
    \label{fig:exp4b}
\end{figure}

\begin{figure}[t]
    \centering
    \includegraphics[scale=0.9]{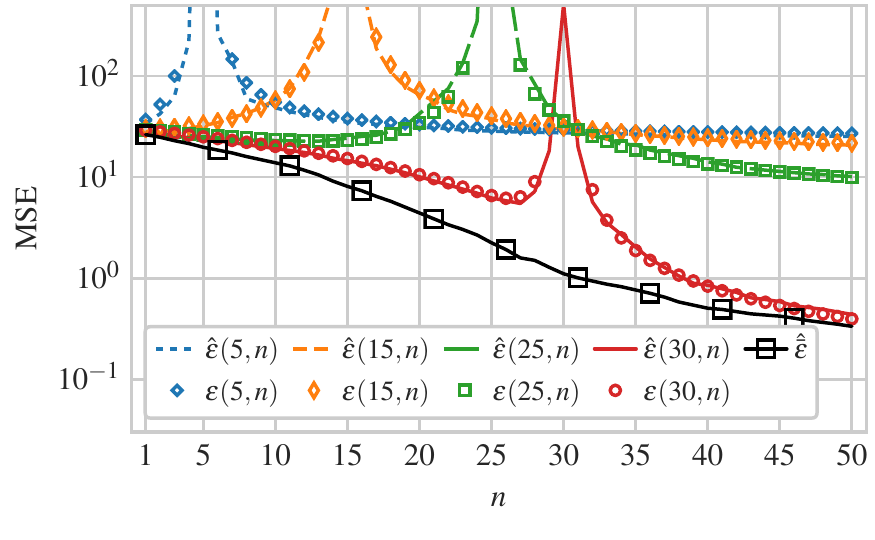}
    \caption{
        The MSE in setting \textit{(S3)}.
        The SNR is high, but the covariance matrix for $\xvec$ is randomized, rather than being identity.
    }
    \label{fig:exp3b}
\end{figure}

\begin{figure}[t]
    \centering
    \includegraphics[scale=0.9]{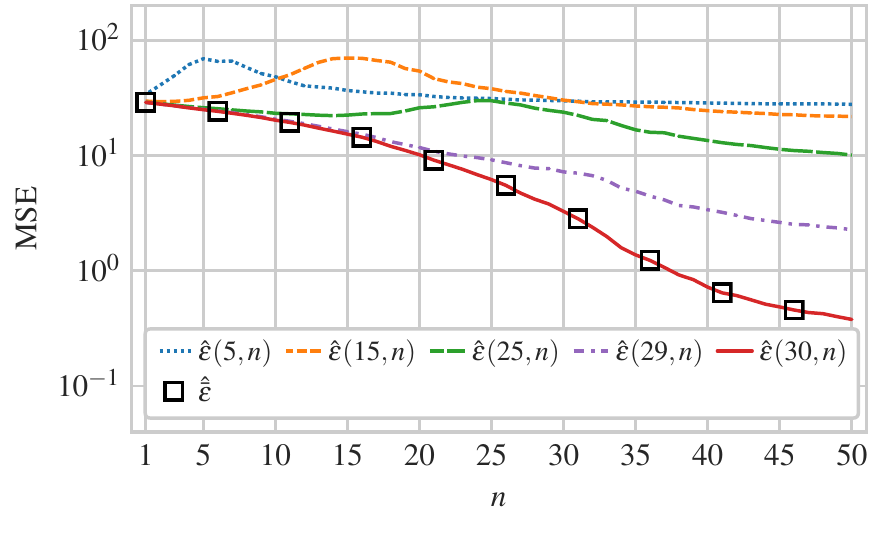}
    \caption{
        The empirical MSE in setting \textit{(S4)}.
        Here the partial LMMSE estimator knows the noise level, i.e., $\sigmahat_z=\sigma_v$.
    }
    \vspace{-1.15em} 
    \label{fig:exp2b}
\end{figure}

\subsection{Experimental Setup}
\kern-0.25em
The numerical results are obtained by  averaging over $M=100$ simulations.
In each simulation $(j)$, $j=1,\dots,M$, 
we draw one random vector $\xvec^{(j)}$ from a Gaussian distribution $\Nc(0,\Kmat_x)$,
one random vector $\vvec^{(j)}$ from $\Nc(0,\Kmat_v)$ and one matrix $\Amat^{(j)}$ where each row is drawn from $\Nc(0,\eye{p})$.
The matrix $\Amat_S^{(j)}$ is extracted as the first $p_S$ columns of $\Amat^{(j)}$.
The partial estimate $\xhat_S^{(j)}$ is then created using $\Wmat_S^{(j)}$ from \eqref{eqn:lmmse_partial}.
The MSE $J^{(j)}(\Wmat_S^{(j)})$ is then computed as 
\begin{equation}
    J^{(j)}(\Wmat_S^{(j)}) = \|\xvec_S^{(j)} - \xhat_S^{(j)}\|^2+\tr(\Kmat_{\xvec_C}),
\end{equation}
and averaged over the $M$ simulations to create the empirical average MSE,
as an estimate of $\egen$: 
\begin{equation}\label{eqn:mse_empirical}
    \hat{\egen}(p_S,n) \triangleq \frac{1}{M} \sum_{i=1}^{M} J^{(j)}(\Wmat_S^{(j)}).
\end{equation}

We set $p=30$ and vary $p_S$ and $n$  to illustrate how $\hat{\egen}$ changes.
We have $\Kmat_{\vvec} = \sigma_v^2 \eye{n}$ and $\Khat_{\zvec}=\sigmahat_z^2 \eye{n}$. 
We consider the following experiment scenarios \textit{(S1)}\:--\:\textit{(S4)} in terms of assumptions on $\Kmat_{\xvec}$, $\Khat_{\xvec_S}$, $\sigma_v$ and $\sigmahat_z$:
    \textit{(S1)} 
        $\Kmat_{\xvec} = \eye{p}$, 
        $\Khat_{\xvec_S} = \eye{p_S}$, 
        $\sigma_v = 0.5$, 
        $\sigmahat_z = 0$;
    \textit{(S2)}
        $\Kmat_{\xvec} = \eye{p}$, 
        $\Khat_{\xvec_S} = \eye{p_S}$, 
        $\sigma_v = \sqrt{p}$, 
        $\sigmahat_z = 0$;
    \textit{(S3)}
        $\Kmat_{\xvec} = \frac{p}{\tr(\Cmat_{\xvec}\T \Cmat_{\xvec})} \Cmat_{\xvec}\T \Cmat_{\xvec}$, with the entries of $\Cmat_{\xvec}\in\Rbb^{p\times p}$ sampled once from $\Nc(0,1)$ and fixed throughout the $M$ simulations,
        $\Khat_{\xvec_S} = \eye{p_S}$,
        $\sigma_v = 0.5$,
        $\sigmahat_z = 0$;
    \textit{(S4)}
        $\Kmat_{\xvec} = \eye{p}$, 
        $\Khat_{\xvec_S} = \eye{p_S}$, 
        $\sigma_v = \sigmahat_z =  0.5$.
Note that in the settings \textit{(S1)}, \textit{(S3)} and \textit{(S4)}, the SNR $10\log_{10}(\tr(\Kmat_{\xvec})/\sigma_v^2)$ is around $21$ dB, 
and in setting \textit{(S2)}, the SNR is at $0$ dB.

\kern-0.5em
\subsection{The MSE, Model Order and the Number of Samples}
\kern-0.5em
In Figures~\ref{fig:exp1b}\:--\:\ref{fig:exp3b} we plot the empirical average MSE $\hat{\egen}$ from \eqref{eqn:mse_empirical} together with the analytical expected MSE $\egen=\egen_S + \tr(\Kmat_{\xvec_C})$ with $\egen_S$ from \eqref{eqn:thm:digaonal:mse} versus the number of samples $n$, 
for the experiments \textit{(S1)}\:--\:\textit{(S3)}.
The empirical values are marked with lines, and the analytical with markers.
We also plot the empirically averaged MSE $\hat{\Bar{\egen}}$ of the true LMMSE estimator in \eqref{eqn:W:lmmse_true_A} as a line with sparsely placed markers.

\textbf{Overview:}
In the plots, we observe a perfect match between the empirical and the analytical curves, confirming our analytical results.
There are clear peaks in MSE when the number of samples $n$ is close to the number of parameters $p_S$ in the assumed model,
as expected from the behaviour of $\gamma$ in \eqref{eqn:thm:digaonal:mse}.
It is clear that in this estimation setting, 
one should avoid having $n$ close to $p_S$, 
and that changing the number of samples or the assumed model order can significantly improve the MSE.

\textbf{Effect of $n$ and $p_s$:}
Consistently over all figures and all $p_S$,
the MSE decreases monotonically as $n$ grows, for $n > p_S$.
However, 
to have performance close to that of the true LMMSE estimator,
$p_S$ must be equal to $p=30$.
For other $p_S$, 
there is a gap in the MSE between the partial and the true estimator which does not vanish as $n$ grows.
In \textit{(S1)}, 
where the model on $\xvec_S$ is correct and the SNR is high,
the MSE with $p_S=p$ is close to that of the true estimator's for all $n$
(except for when $n$ is close to $p_S$),
despite the fact that the assumed model ignores the noise.
Furthermore, even in \textit{(S2)} and \textit{(S3)}, 
where the SNR is either low (and noise is still ignored in the assumed estimator), 
or the assumed model on $\xvec_S$ is incorrect, it is still possible to get error values comparable to that of the true estimator with $p_S=p$  for large $n$. 

\textbf{Effect of SNR:}
Figure~\ref{fig:exp4b}  illustrates the result of Corollary~\ref{col:n_SNR_dependency},
i.e., if $n$ is large enough $n > p + \sigma_v^2/\sigma_x^2 + 1$,
then the MSE decreases monotonically with $p_S$.
This bound on $n$ provides a clear marking of an operating range on which performance gain is guaranteed when increasing $p_S$.
In this low SNR setting,
there is a large range of $n$ on which a smaller $p_S$ gives lower MSE.
In other words,
although we have the correct model on $x_S$, 
choosing a smaller model size $p_S$ can improve the performance even after interpolation threshold (i.e., $n=p$),
if the estimator ignores the noise ($\sigmahat_z = 0$) and there is not enough data.

\textbf{Matched noise level:} 
In Figure~\ref{fig:exp2b}, we plot the results of experiment \textit{(S4)},
where the assumed noise level is the same as the level of the true noise: $\sigmahat_z = \sigma_v$.
There are still peaks in the MSE but they are significantly dampened compared to the earlier experiments.
For high values of $p_S$,
e.g., $p_S=29$, $30$, the MSE decreases monotonically with $n$.
For lower values of $p_S$, we still have the same effects on the MSE as we saw in the settings \textit{(S1)}\:--\:\textit{(S3)}.
More specifically, 
as in previous experiments,
the performance is not guaranteed to improve with increasing $n$ or $p_S$.

%% file: conclusions.tex
\vspace*{-3pt}
\section{Conclusions}\label{sec:conclusions}
\vspace*{-3pt}
Under an LMMSE estimation framework, 
we investigated the average degradation of the estimation performance due to model mismatch. 
Our analytical results,
verified with simulations, illustrate the interplay between the SNR, the number of samples and the model orders of the underlying and the assumed models. In general, neither increasing the number of samples nor the assumed model complexity can guarantee performance improvement.
Extending these results to different covariance structures on the regressors as well as to complex regressors, including non-circular scenarios,
are important directions for future work.

%% file: appendix/appendix.tex
\setlength{\abovedisplayskip}{2pt}
\setlength{\belowdisplayskip}{2pt}
\section{Appendix}
\vspace*{-5pt}
\subsection{Proof of Theorem~\ref{thm:mse}}\label{proof:thm:mse}
\vspace*{-2pt}
In the setting of Theorem~\ref{thm:mse},
the partial LMMSE estimator in \eqref{eqn:lmmse_partial} is
$
    \Wmat_S = \Amat_S\T(\Amat_S \Amat_S\T)\p = \Amat_S\p.
$
Plugging this $\Wmat_S$ into \eqref{eqn:MSE_trace_xhat_S},
and applying the trace operator,
we have
\begin{align}\begin{split}
   \!\! J_S(\Amat_S\p) 
    = & \tr\big( (\eye{p_S} - \Amat_S\p \Amat_S)\T (\eye{p_S} - \Amat_S\p \Amat_S) \Kmat_{\xvec_S} \\
        & + \Amat_C\T (\Amat_S\p)\T \Amat_S\p \Amat_C \Kmat_{\xvec_C} + (\Amat_S\p)\T \Amat_S\p \Kmat_{\vvec} \\
        & - 2(\eye{p_S} - \Amat_S\p \Amat_S)\T \Amat_S\p \Amat_C \Kmat_{\xvec_C \xvec_S}\big),
\end{split}
\end{align}
By the definition of the pseudoinverse,
the matrix $\Amat_S\p \Amat_S$ is symmetric,
$\Amat_S\p \Amat_S \Amat_S\p = \Amat_S\p$.
Hence,  
$
    (\eye{p_S} - \Amat_S\p \Amat_S)\T (\eye{p_S} - \Amat_S\p \Amat_S) = (\eye{p_S} - \Amat_S\p \Amat_S),
$
and
$
(\eye{p_S} - \Amat_S\p \Amat_S)\T \Amat_S\p = \Amat_S\p - \Amat_S\p = 0.
$
Furthermore, the pseudoinverse has the property 
$(\Amat_S\p)\T \Amat_S\p = (\Amat_S \Amat_S\T)\p$.
We can now write
\begin{align}\begin{split}
    \! J_S(\Amat_S\p) 
    = & \tr\big( \Kmat_{\xvec_S} - \Amat_S\p \Amat_S \Kmat_{\xvec_S} \\
        & + \! \Amat_C\T (\Amat_S \Amat_S\T)\p \Amat_C \Kmat_{\xvec_C} \! + \! (\Amat_S \Amat_S\T) \p \Kmat_{\vvec}\big).
\end{split}
\end{align}
We now take the expectation over the distribution of $\Amat$, noting that $\Amat_S$ and $\Amat_C$ are uncorrelated,
and use the linearity and cyclic property of the trace operator to write:
\begin{align}
\begin{split}
    \egen(p_S,n) = \tr(\Kmat_{\xvec_S}) - \tr\left(\eunder{\Amat_S}\left[\Amat_S\p \Amat_S \right] \Kmat_{\xvec_S}\right) \\
    +\tr\left( \eunder{\Amat_S}\left[ (\Amat_S \Amat_S\T)\p \right] \eunder{\Amat_C}\left[ \Amat_C \Kmat_{\xvec_C} \Amat_C\T \right] \right) \\
    +\tr\left(\eunder{\Amat_S}\left[ (\Amat_S \Amat_S\T)\p \right] \Kmat_{\vvec} \right).
\end{split}
\end{align}

We continue the proof by noting that 
\begin{align}
    \tr\left( \eunder{\Amat_S}[ \Amat_S\p \Amat_S] \Kmat_{\xvec_S} \right) 
    & = \eunder{\xvec_S}\left[\xvec_S\T \eunder{\Amat_S} [ \Amat_S\p \Amat_S ] \xvec_S\right] \\
    & = \tfrac{\min\{p_S, n\}}{p_S} \tr(\Kmat_{\xvec_S}),
\end{align}
where in the last step we used Lemma~3 of \cite{HellkvistOzcelikkaleAhlen_distributed2020_spawc},
as well as $E_{\xvec_S}[\|\xvec_S\|^2] = \tr(\Kmat_{\xvec_S})$.
By \cite{cook_forzani_wishart_2011}, 
we have that $\Ebb_{\Amat_S}[(\Amat_S \Amat_S\T)\p] = \frac{1}{n} \gamma \eye{n}$, 
with $\gamma$ as in \eqref{eqn:gamma}.
We can now write
\begin{align}
\begin{split}
    &\egen_S(p_S,n) = \tr( \Kmat_{\xvec_S}) - \tfrac{\min\{p_S, n\}}{p_S} \tr(\Kmat_{\xvec_S}) \\
    &\quad \quad  +\frac{1}{n} \gamma \tr\left( \Kmat_{\xvec_C} \eunder{\Amat_C}\left[ \Amat_C\T \Amat_C \right] \right)
    + \frac{1}{n} \gamma \tr(\Kmat_{\vvec}),
\end{split}
\end{align}
into which we plug in that 
$
    \Ebb_{\Amat_C}[ \Amat_C\T \Amat_C ]  = n \eye{p_C},
$
to yield the final expression for $\egen_S(p_S, n)$.